\documentclass{article}
\usepackage[T1]{fontenc} 
\usepackage[utf8]{inputenc} 

\def\authorname{A. S\'{a} Pinto, I. Domingues, and M. E. P. Davies}

\def\papertitle{Shift If You Can: Counting and Visualising Correction Operations for Beat Tracking Evaluation}

\usepackage{ismir-lbd}

\usepackage{lineno}

\title{Shift If You Can: Counting and Visualising Correction Operations for Beat Tracking Evaluation}

\multauthor
{Ant\'{o}nio S\'{a} Pinto$^{1}$ \hspace{1cm}  
In\^{e}s Domingues$^{2}$ \hspace{1cm} Matthew E. P. Davies$^{3}$} { 
$^1$ INESC TEC and University of Porto, Faculty of Engineering, Portugal\\
$^2$ IPO Porto Research Centre (CI-IPOP), Portugal\\
$^3$ University of Coimbra, CISUC, DEI, Portugal\\
{\tt antonio.s.pinto@inesctec.pt}
}
\def\authorname{A. S\'{a} Pinto, I. Domingues, and M. E. P. Davies}

\begin{document}

\maketitle
\begin{abstract}
In this late-breaking abstract we propose a modified approach for beat tracking evaluation which poses the problem in terms of the effort required to transform a sequence of beat detections such that they maximise the well-known F-measure calculation when compared to a sequence of ground truth annotations. Central to our approach is the inclusion of a shifting operation conducted over an additional, larger, tolerance window, which can substitute the combination of insertions and deletions. We describe a straightforward calculation of annotation efficiency and combine this with an informative visualisation which can be of use for the qualitative evaluation of beat tracking systems. We make our implementation and visualisation code freely available in a GitHub repository.
\end{abstract}

\section{Introduction}


In musically creative tasks which rely on some prior analysis of a music signal, (e.g., musical audio beat tracking for remixing or beat-matching) the mean performance of an algorithm over large datasets is of less relevance than knowing its exact performance for a given piece. Indeed, within Creative MIR, high importance is placed on the extremely accurate analysis of music signals \cite{andersen2016ismir}. Errors in analysis can have an immediate negative impact on the musical result, as well as being a confounding factor when seeking to evaluate the core creative task. To mitigate this problem, we have two options: to exploit high-level user-input to guide how the analysis is undertaken \cite{sapinto2019cmmr} towards a very accurate analysis from the outset; and/or to manually correct the output of the algorithm using a tool such as Sonic Visualiser \cite{cannam2010acm} until it is satisfactory for the end-user. To reduce the workload and cognitive burden, we normally wish to minimize the number of user interactions required.

For musical audio beat tracking, the alteration of beat detections may be challenging due to the underlying difficulty of the musical material, but the correction process can be achieved using two simple editing operations: insertions and deletions -- combined with repeated listening to audible clicks mixed with the input. The number of insertions and deletions correspond to counts of \textit{false negatives} and \textit{false positives} respectively, and form part of the calculation of the well-known F-measure. While the F-measure is routinely used in beat tracking (and many other MIR tasks) to measure accuracy, we can also view it in terms of the 
effort required to transform an initial set of beat detections to a final desired result (e.g., a ground truth annotation sequence). In this way, a high F-measure would imply low effort in manual correction and vice-versa. 

In practice, correcting beat detections often relies on a third operation: the \textit{shifting} of poorly localised individual beats. This shifting operation is particularly relevant when correcting tapped beats, which can be subject to human motor noise as well as jitter and latency during acquisition. Under the logic of the F-measure calculation, shifting beat detections which fall outside tolerance windows are effectively counted twice: as a false positive \textit{and} a false negative. We argue that for beat tracking evaluation this creates a modest, but important, disconnect between common practice in annotation correction and a widely-used evaluation method. We propose that, wherever reasonable, the single operation of shifting should be prioritised over a deletion followed by an insertion. 

\section{Measuring Annotation Efficiency}

We devise a straightforward calculation for the \textit{annotation efficiency} based on counting the number of shifts, insertions, and deletions. We provide an open-source Python implementation\footnote{\url{https://github.com/MR-T77/ShiftIfYouCan}} which graphically displays the minimum set and type of operations required to transform a sequence of initial beat detections in such a way as to maximize the F-measure when comparing the transformed detections against a ground truth annotation sequence. It is important to note that our goal is \textit{not} to transform the beat detections such that they are absolutely identical to the ground truth, (although such transformations are theoretically possible) but rather to perform as few operations as possible to ensure $F=1.00$, subject to a user-defined tolerance window.



\begin{figure*}[h!]
\centering
  \includegraphics[width=17cm]{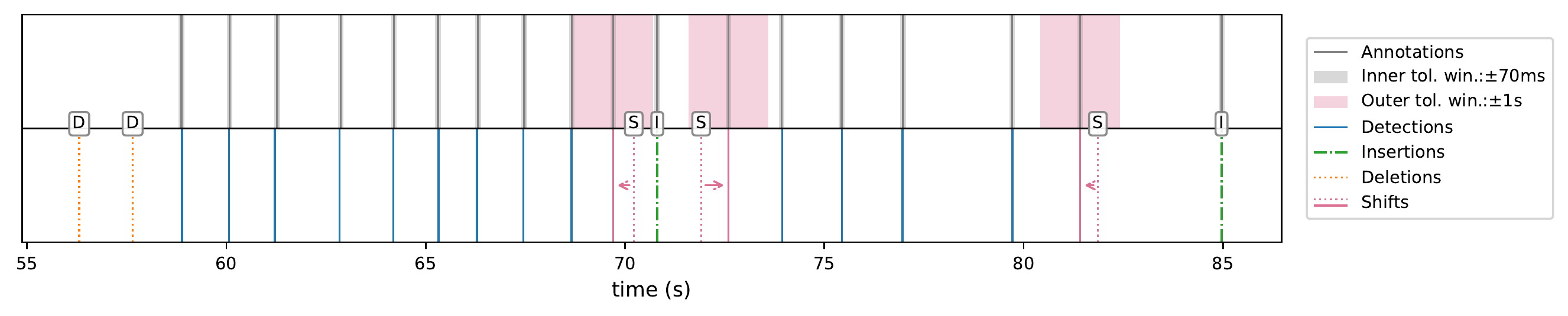}
  \includegraphics[width=17cm]{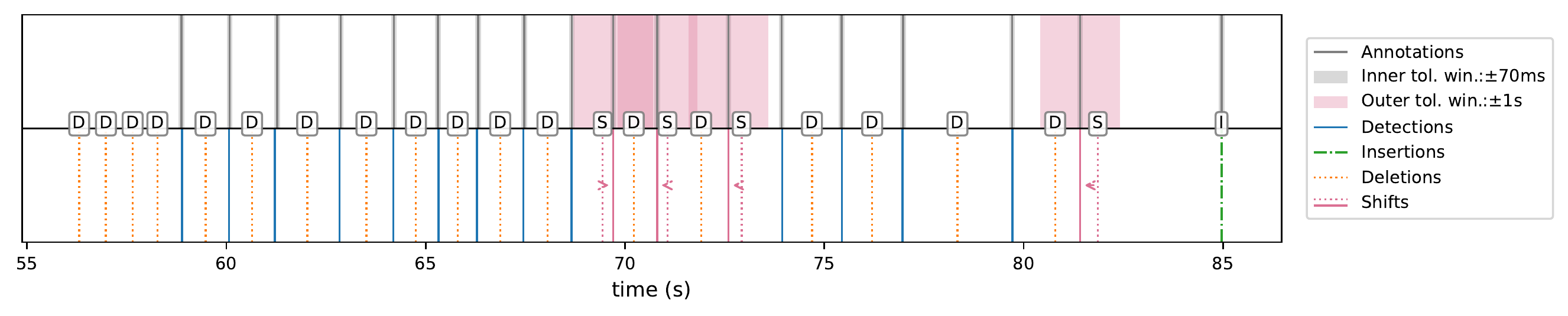}
  \caption{Visualisation of the operations required to transform beat detections to maximize the F-measure when compared to the ground truth annotations. (Top) Original beat detections vs. ground truth annotations. (Bottom) Double variation of beat detections vs. ground truth annotations. The inner tolerance window is overlaid on all annotations, where as the outer tolerance window is only shown for those detections to be shifted.}
  \label{fig:regular}
\end{figure*}


Around each ground truth annotation, we create an \textit{inner tolerance window} (set to $\pm70$\,ms) and count the number of \textit{true positives} (unique detections), $t^{+}$. We mark each matching detection and annotation pair as ``accounted for'' and remove them from further analysis. All remaining detections then become candidates for \textit{shifting} or \textit{deletion}. For each remaining annotation, we then look for the closest unaccounted for detection within an \textit{outer tolerance window} (set to $\pm1$\,s), which we use to reflect a localised working area for manual correction. If any detection exists, we mark it as a shift along with the required temporal correction offset. After the analysis of all unaccounted for annotations is complete, we count the number of shifts, $s$. Any remaining annotations correspond to false negatives, $f^{-}$, with leftover detections marked for deletion and counted as false positives, $f^{+}$. To give a measure of annotation efficiency we adapt the evaluation method in \cite{dixon2001jnmr} to include the shifts: 
\begin{equation}
ae = t^{+} / (t^{+} + s + f^{+} + f^{-}).  
\end{equation}
Reducing the inner tolerance window transforms true positives into shifts and thus sends $t$ and hence $ae$ to $0$. In this limit, the modified detections are then identical to the target sequence.

To allow for metrical ambiguity in beat tracking evaluation, it is common to create a set of variations of the ground truth by interpolation and sub-sampling operations. In our implementation, we flip this behaviour, and instead create variations of the detections. In this way, we can couple a global operation applied to all detections (e.g., interpolating all detections by two), with the subsequent set of local correction operations. Whichever variation has the highest annotation efficiency represents the shortest path to obtaining an output consistent with the annotations.   

The fundamental difference of our approach compared to the standard F-measure is that we view the evaluation from a user workflow perspective, and essentially \textbf{we shift if we can}. By recording each individual operation, we can count them for evaluation purposes as well as visualising them, as shown in Figure.~\ref{fig:regular} which contrasts the use of the original beat detections compared to the double variation of the beats. The example shown is from the composition ``Evocaci\`{o}n'' by Jose Luis Merlin. It is a solo piece for classical guitar which features extensive \textit{rubato} and is among the more challenging pieces in the Hainsworth dataset \cite{Hainsworth2004}. By inspection we can see the original detections are much closer to the ground truth than the double variation. They require just two deletions (covering a pause in the performance), three shifts, and two insertions, and thus the annotation efficiency is much higher: $0.650$ compared to $0.361$.

The precise recording of the set of individual operations allows additional deeper evaluation which can indicate precisely which operations are most beneficial and in which order. For the F-measure, shifts are always more beneficial than the isolated insertions or deletions, but for other evaluation methods, i.e., those which measure continuity, the temporal location of the operation may be more critical. By viewing evaluation from a transformation perspective combined with an informative visualisation, we hope our implementation can contribute to a better qualitative understanding of beat tracking algorithms. 

Looking beyond beat tracking, our approach can readily be applied to other temporal MIR tasks, e.g., onset detection and structural boundary detection -- subject to appropriate tolerance windows. In future work, we intend to expand the functionality to additionally correct labelled time instants, e.g., beat positions in each bar when estimating downbeats. We will also look to position our work in the context of the formal theory of edit distances. Finally, we plan to undertake a user-study to understand how well our measure of annotation efficiency correlates with user-reported effort in the manual correction of beat estimates.

\section{Acknowledgments}
This work is funded by national funds through the FCT - Foundation for Science and Technology, I.P., within the scope of the project CISUC - UID/CEC/00326/2020 and by European Social Fund, through the Regional Operational Program Centro 2020. It is also supported by the FCT-Foundation for Science and Technology, I.P., under the grant SFRH/BD/120383/2016 and the projects IF/01566/2015 and UIDP/00776/2020.
\bibliography{ISMIR2020lbd}

%
%
%
%
%

\end{document}